\documentstyle[12pt,fleqn,epsf,epsfig,axodraw]{article}
\def\lsim{\raise0.3ex\hbox{$\;<$\kern-0.75em\raise-1.1ex\hbox{$\sim\;$}}}
\def\gsim{\raise0.3ex\hbox{$\;>$\kern-0.75em\raise-1.1ex\hbox{$\sim\;$}}}

\begin{document}
\setlength{\unitlength}{1cm}
\setlength{\mathindent}{0cm}
\thispagestyle{empty}
\null
\hfill WUE-ITP-03-013\\
\null
\hfill UWThPh-2003-18\\
\null
\hfill HEPHY-PUB 776/03\\
\null
\hfill LC-TH-2003-065\\
\null
\hfill hep-ph/0308143\\
\vskip .8cm
\begin{center}
{\Large \bf CP asymmetries in $ e^+e^- \to\tilde{\chi}^0_i \tilde{\chi}^0_j$}
\vskip 2.5em
{\large
{\sc A.~Bartl$^{a}$\footnote{e-mail:
        bartl@ap.univie.ac.at}
     H.~Fraas$^{b}$\footnote{e-mail:
        fraas@physik.uni-wuerzburg.de},
	  O.~Kittel$^{a,b}$\footnote{e-mail:
		  kittel@physik.uni-wuerzburg.de},
	  W.~Majerotto$^{c}$\footnote{e-mail:
        majer@qhepu3.oeaw.ac.at}
}}\\[1ex]
{\normalsize \it
$^{a}$ Institut f\"ur Theoretische Physik, Universit\"at Wien, 
Boltzmanngasse 5, A-1090 Wien, Austria}\\
{\normalsize \it
$^{b}$ Institut f\"ur Theoretische Physik, Universit\"at
W\"urzburg, Am Hubland, D-97074~W\"urzburg, Germany}\\
{\normalsize \it
$^{c}$ Institut f\"ur Hochenergiephysik, \"Osterreichische
Akademie der Wissenschaften, Nikolsdorfergasse 18, 
A-1050 Wien, Austria}\\
\vskip 1em
\end{center} \par
\vskip .8cm

\begin{abstract}

We study two CP sensitive triple-product asymmetries for
neutralino production 
$e^+e^- \to\tilde{\chi}^0_i \tilde{\chi}^0_j$
and the subsequent leptonic two-body decays 
$\tilde{\chi}^0_i \to \tilde{\ell} \, \ell$ of one of the neutralinos
and of the slepton
$ \tilde{\ell} \to \tilde{\chi}^0_1 \, \ell$ 
at an $e^+e^-$-linear collider with $\sqrt{s}=500$ GeV.
We calculate the asymmetries in the Minimal Supersymmetric 
Standard Model with complex parameters $\mu$ and $M_1$.
We show that the largest asymmetries are 10$\%$ or 25$\%$ 
and estimate the event rates which are necessary to measure 
the asymmetries.
Polarized electron and positron beams can significantly enhance the 
asymmetries and cross sections.
In addition, we show how the two decay leptons
can be distinguished by making use of their 
energy distributions.

\end{abstract}

\newpage

\section{Introduction}

The Minimal Supersymmetric Standard Model (MSSM) \cite{haberkane} 
contains new sources of CP violation if the parameters of the model
are complex. 
In the neutralino sector of the MSSM these are the $U(1)$ and $SU(2)$
gaugino mass parameters $M_1$ and $M_2$, respectively, and the higgsino mass
parameter $\mu$. One of these parameters, usually $M_2$, can be made 
real by redefining the fields. 
The non-vanishing phases of $M_1$ and $\mu$ cause CP-violating effects 
already at tree level, which could be large and thus 
observable in high energy collider experiments \cite{TDR}.

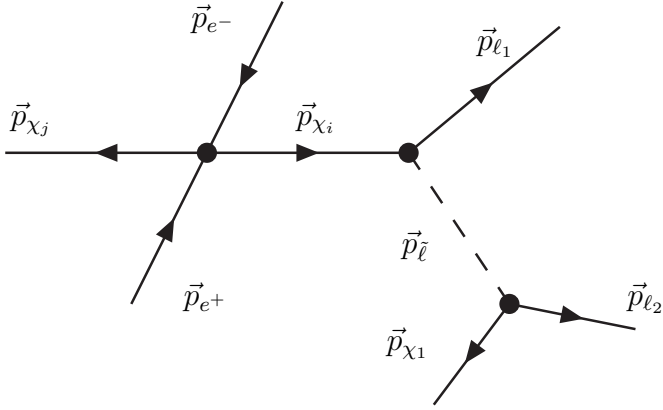
\begin{figure}[h]
\begin{picture}(5,6.)(-2,.5)
		\put(1,4.7){$\vec p_{\chi_j}$}
   \put(3.4,6){$\vec p_{e^- }$}
   \put(3.3,2.3){$\vec  p_{e^+}$}
   \put(4.8,4.7){$\vec p_{\chi_i}$}
   \put(7.2,5.7){$ \vec p_{\ell_1}$}
   \put(6.2,3.){$ \vec p_{\tilde{ \ell} }$}
   \put(9.2,2.3){$\vec  p_{\ell_2}$}
   \put(6,1.7){$ \vec p_{\chi_1}$}
\end{picture}
\scalebox{1.9}{
\begin{picture}(0,0)(1.3,-0.25)
\ArrowLine(40,50)(0,50)
\Vertex(40,50){2}
\ArrowLine(55,80)(40,50)
\ArrowLine(25,20)(40,50)
\ArrowLine(40,50)(80,50)
\ArrowLine(80,50)(110,75)
\DashLine(80,50)(100,20){4}
\Vertex(80,50){2}
\ArrowLine(100,20)(125,15)
\ArrowLine(100,20)(85,0)
\Vertex(100,20){2}
\end{picture}}
\caption{\label{shematic picture}
          Schematic picture of the neutralino production
          and decay process.}
\end{figure}

In this note we study neutralino production
(for recent studies with complex parameters and polarized beams see
\cite{choi1,kali,gudi1} )
\begin{eqnarray} \label{production}
   e^-+e^+&\to&
	\tilde{\chi}^0_i+\tilde{\chi}^0_j 
\end{eqnarray}
and the subsequent leptonic two-body decay of one neutralino
\begin{eqnarray} \label{decay_1}
   \tilde{\chi}^0_i&\to& \tilde{\ell} + \ell_1,  
\end{eqnarray}
and of the decay slepton
	\begin{eqnarray} \label{decay_2}
  \tilde{\ell}&\to&\tilde{\chi}^0_1+ \ell_2;\;\;\; \ell= e,\mu.
\end{eqnarray}
The decay of  the other neutralino $\tilde{\chi}^0_j$
is not considered. For a schematic picture of the production and
decay process see Fig.~\ref{shematic picture}.

T-odd observables \cite{choi1,donoghue,oshimo,valencia,staudecay} 
are a useful tool to study the CP-violating effects of the 
parameters $M_1$ and $\mu$. 
They involve triple-products of particle momenta or spin vectors.
For neutralino production (\ref{production}) and  the 
two-body decay chain of the neutralino (\ref{decay_1})-(\ref{decay_2})
we introduce the triple-product 
 \begin{eqnarray}\label{AT1}
	 {\mathcal T}_{I} &=& (\vec p_{e^-} \times \vec p_{\chi_i}) 
	 \cdot \vec p_{\ell_1},
 \end{eqnarray}
which changes sign under time reversal. We define the T-odd asymmetry 
 \begin{eqnarray}
	 {\mathcal A}_{I} = \frac{\sigma({\mathcal T}_{I}>0)
						 -\sigma({\mathcal T}_{I}<0)}
							{\sigma({\mathcal T}_{I}>0)+
								\sigma({\mathcal T}_{I}<0)}
							=\frac{N_+ - N_- }{N_+ + N_-},
\label{TasymmetryI}
\end{eqnarray}
where $\sigma$ is the cross section 
$\sigma(e^-e^+\to\tilde\chi^0_i\tilde\chi^0_j ) \times
{\rm BR}(\tilde \chi^0_i\to\tilde\ell\ell_1)\times
{\rm BR}(\tilde\ell\to\tilde\chi^0_1\ell_2)$ defined below.
The asymmetry ${\mathcal A}_{I}$ is thus the difference between 
the number of events with the lepton $\ell_1$ above $(N_+)$ and below  
$(N_-)$ the production plane. 

One can show that ${\mathcal A}_{I}$ is due to the 
polarization of the neutralino perpendicular to the production
plane, which is non-vanishing only if there are CP-violating 
phases in the neutralino sector and if different neutralinos 
are produced \cite{choi1,kali,gudi1}.
In order to measure the asymmetry ${\mathcal A}_{I}$, the 
momentum  $\vec p_{\chi_i}$ of neutralino $\tilde{\chi}^0_i$ 
and thus the production plane has to be reconstructed. 

Considering also the leptonic two-body decay of the slepton (\ref{decay_2}),
we can construct another T-odd observable replacing the 
neutralino momentum $\vec p_{\chi_i}$ in Eq.~(\ref{AT1}) by the lepton momentum
$\vec p_{\ell_2}$ from the slepton decay. We define the second
triple-product and the corresponding T-odd asymmetry as
\begin{eqnarray}
	{\mathcal T}_{II} = (\vec p_{e^-} \times 
		\vec p_{\ell_2})\cdot \vec p_{\ell_1}, \quad \quad
	{\mathcal A}_{II} = \frac{\sigma({\mathcal T}_{II}>0)
						 -\sigma({\mathcal T}_{II}<0)}
							{\sigma({\mathcal T}_{II}>0)+
							\sigma({\mathcal T}_{II}<0)}.
	\label{AT2}
\end{eqnarray}
Measuring ${\mathcal A}_{II}$ does not require the reconstruction of
the neutralino momentum. Like in ${\mathcal A}_{I}$, the contributions 
to ${\mathcal A}_{II}$ also stem from the CP-violating neutralino polarization.
However, ${\mathcal A}_{II}$ is smaller than ${\mathcal A}_{I}$
because the CP-violating contribution from the production is 
washed out by the kinematics of the slepton decay (\ref{decay_2}).
Due to CPT invariance, the T-odd asymmetries (\ref{TasymmetryI})
and (\ref{AT2}) are CP-odd 
if the widths of the exchanged particles and final state
interactions are neglected, which is done in this work.

The cross section in the laboratory system is obtained by integrating 
the amplitude squared $|T|^2$ over the Lorentz invariant phase space element 
$d{\rm Lips}(s,p_{\chi_j },p_{{\ell}_1},p_{\chi_1},p_{{\ell}_2})$:
\begin{equation}\label{crossection}
		d\sigma=\frac{1}{2 s}|T|^2 d{\rm Lips}; \quad \quad 
 |T|^2 = 4~|\Delta(\tilde{\chi}^{0}_i)|^2~  
		|\Delta(\tilde{\ell})|^2
	  ( P D_1 + \vec \Sigma_P \vec\Sigma_{D_1} )\;D_2.
\end{equation}
To the total cross section, only the terms $P D_1$ contribute and
the spin correlation terms $\vec \Sigma_P \vec\Sigma_{D1} $ vanish.
However, the spin correlation terms contribute to the asymmetries 
${\mathcal A}_{I}$ and ${\mathcal A}_{II}$.
The analytical formulae for $P$ and $\Sigma^{a}_P$ are given 
in \cite{gudi1}. The expressions for $D_1,D_2$ and $\Sigma^{a}_{D_1}$
will be given in \cite{ourwork}, where also more details of the
calculation and additional numerical results can be found.
We use the narrow width approximation for the propagators
$ \Delta(k)=1/[s_k-m_k^2+im_k\Gamma_k]$, 
$k=\tilde{\chi}^{0}_i,\tilde{\ell}$.

In order to measure the asymmetries ${\mathcal A}_{I}$ and 
${\mathcal A}_{II}$, the two leptons $\ell_1$ and $\ell_2$ from  the 
neutralino (\ref{decay_1}) and selectron decay (\ref{decay_2}),
respectively, have to be distinguished.
This can be achieved by measuring the energies of the two leptons
and making use of their different energy distributions. 
Owing to the Majorana property of the 
neutralinos, the contributions of the spin correlations to the energy 
distributions vanish if CP is conserved \cite{pectovgudi}.
In the case of CP violation, they vanish only to leading order perturbation
theory \cite{pectovgudi}. In our case, the contributions are
proportional to the widths of the exchanged particles
in the production process (\ref{production}) and thus can be
neglected.

\section{Numerical results
	\label{Numerical results}}

We now present numerical results for 
$e^-e^+\to\tilde\chi^0_1 \tilde\chi^0_2$
with the subsequent decay of $ \tilde\chi^0_2$
into the right selectron and right smuon, 
$ \chi^0_2\to\tilde\ell_R\ell_1$.
We study the dependence of the asymmetries 
and the production cross section on the parameters
$\mu = |\mu| \, e^{ i\,\varphi_{\mu}}$, 
$M_1 = |M_1| \, e^{ i\,\varphi_{M_1}}$ and $M_2$
for $\sqrt{s} = 500$ and longitudinally polarized beams with 
$P_-=0.8$ and $P_+=-0.6$. We fix $\tan \beta=10$ and assume 
$|M_1|=5/3 \tan^2\theta_W M_2$. We use the
renormalization group equations \cite{hall} for the 
selectron and smuon masses,
$m_{\tilde\ell_R}^2 = m_0^2 +0.23 M_2^2
-m_Z^2\cos 2 \beta \sin^2 \theta_W$, taking $m_0=100$ GeV.

\begin{figure}[h]
\begin{picture}(10,9)(0,9)
   \put(0,9){\includegraphics{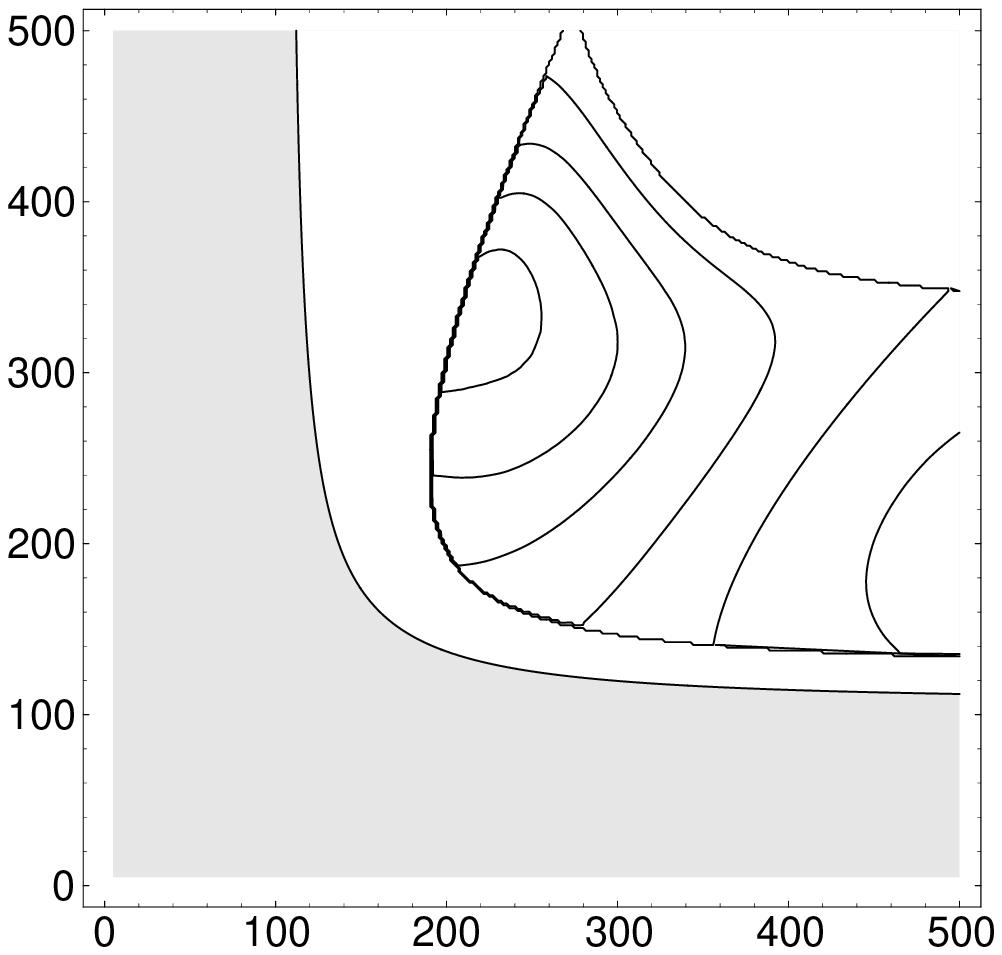}}
	\put(3.5,16.5){\fbox{${\mathcal A}_{II}$ in \% }}
	\put(5.5,8.7){$|\mu|$~/GeV}
	\put(0,16.3){$M_2$~/GeV }
	\put(3.3,13.5){$ 9.5 $}
	\put(3.7,12.9){$ 8 $}
	\put(4.,12.5){$6 $}
	\put(4.4,12.2){$3 $}
  \put(5,12){$0 $}
 \put(6.3,11.8){$-3 $}
   \put(0.5,8.7){Fig.~\ref{plots_12}a}
	\put(8,9){\includegraphics{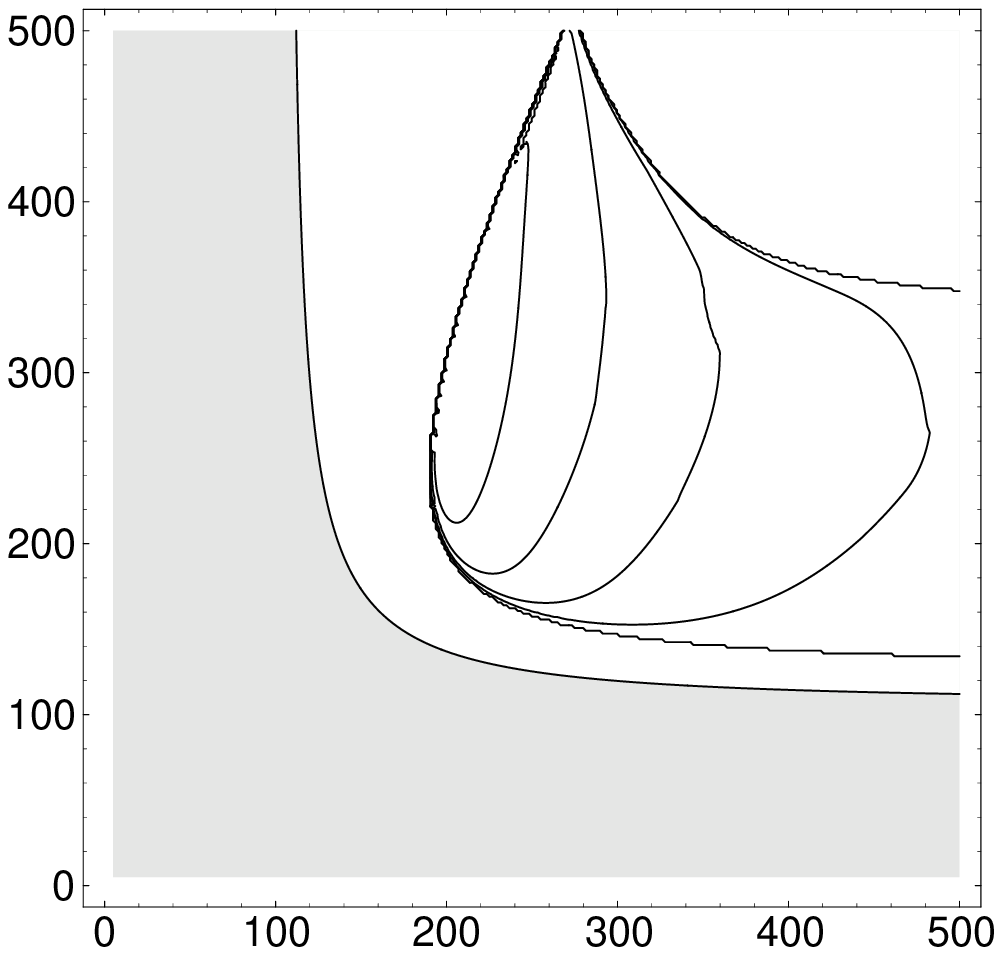}}
	\put(9.9,16.5){\fbox{$\sigma(e^+\,e^- \to\tilde{\chi}^0_1 \,
			\tilde{\chi}^0_1 \ell_1 \,\ell_2 )$ in fb}}
	\put(13.5,8.7){$|\mu|$~/GeV}
	\put(8,16.3){$M_2$~/GeV }
	\put(11.2,13.2){$60$}
   \put(11.7,12.8){$20$}
   \put(12.7,12.7){$4 $}
	\put(13.8,12.5){$0.4 $}
	\put(8.5,8.7){Fig.~\ref{plots_12}b}
\end{picture}
\vspace*{.5cm}
\caption{
	Contour plots for  
	$e^+\,e^- \to\tilde{\chi}^0_1 \, \tilde{\chi}^0_2$;
$\tilde{\chi}^0_2 \to \tilde{\ell}_R \, \ell_1$;
$ \tilde{\ell}_R \to \tilde{\chi}^0_1 \, \ell_2$ for
$ \ell= e,\mu$ 
in the $|\mu|$--$M_2$ plane with $\varphi_{M_1}=0.5\pi $, 
$\varphi_{\mu}=0$, $\tan \beta=10$,  $P_-=0.8$ and $P_+=-0.6$.
\label{plots_12}}
\end{figure}

Fig.~\ref{plots_12}a shows the $|\mu|$--$M_2$ dependence 
of the asymmetry ${\mathcal A}_{II}$ for $\varphi_{M_1}=0.5~\pi $ and $\varphi_{\mu}=0$.
The gray shaded area  is excluded for 
chargino masses $m_{\tilde\chi_1^{\pm}}<104 $ GeV. 
In the blank areas either  the sum of the neutralino masses
is larger than $\sqrt{s}=500$ GeV or the two body decay (\ref{decay_1})
of the neutralino into the right selectron and smuon is not open.

In the region $|\mu|  \lsim 250 $ GeV, where the right selectron
exchange dominates in the neutralino production process, 
the asymmetry reaches large values up to $9.5\%$ for our choice of beam 
polarization. This enhances the asymmetry up to a factor of 2 compared to the
case of unpolarized beams. With increasing  $|\mu|$ the asymmetry
decreases and finally changes sign. This is due to the increasing
contributions of the left selectron exchange which contributes to the 
asymmetry with opposite sign and dominates for $|\mu| \gsim 300$ GeV
because of the larger $\tilde\chi^0_2-\tilde e_L$ coupling.
In this region the asymmetry can be enhanced up to a factor 2 
by reversing the signs of both beam polarizations.

The contour lines of the cross section
$\sigma(e^-e^+\to\tilde\chi^0_1\tilde\chi^0_2 ) \times
{\rm BR}(\tilde \chi^0_2\to\tilde\ell_R\ell_1)\times
{\rm BR}(\tilde\ell_R\to\tilde\chi^0_1\ell_2)$
with BR($ \tilde\ell_R \to \tilde\chi^0_1\ell_2$) = 1
is shown in Fig.~\ref{plots_12}b in the $|\mu|$--$M_2$ plane 
for $\varphi_{\mu}=0$ and $\varphi_{M_1}=0.5~\pi$. 
For $|\mu| \lsim 250 $ GeV the right selectron exchange dominates
in the production process $e^-e^+\to\tilde\chi^0_1\tilde\chi^0_2 $
and is enhanced by our choice of beam polarization $P_-=0.8$ and
$P_+=-0.6$. No beam polarization would reduce the values of
the cross section in that region by a factor as small as 0.4. 
For  $|\mu| \gsim 300$ GeV, a sign reversal of both polarizations 
would enhance the cross section by a factor between 1 and 20.

The branching ratio ${\rm BR} (\tilde\chi^0_2 \to \tilde\ell_R\ell )$
(summed over both signs of charge) is more than 40\% in those regions
where the cross section is larger than 20 fb.
The branching ratio decreases with increasing $|\mu|\gsim 300$ GeV if the 
two-body decays into the lightest neutral Higgs boson $h^0$  and/or the
$Z$ boson are kinematically allowed. The decay into left selectrons and smuons 
can be neglected  because these channels are either not open or the branching 
fraction is smaller than 1\%. For $M_2 \lsim 200$ GeV the decay into the 
lightest stau $\tilde \tau_1$ is larger than 80\% if, for example,
$A_{\tau} = -250$ GeV.

\begin{figure}[h]
			\begin{minipage}{0.45\textwidth}
 \begin{picture}(10,8)(0,0)
	\put(0,0){\includegraphics{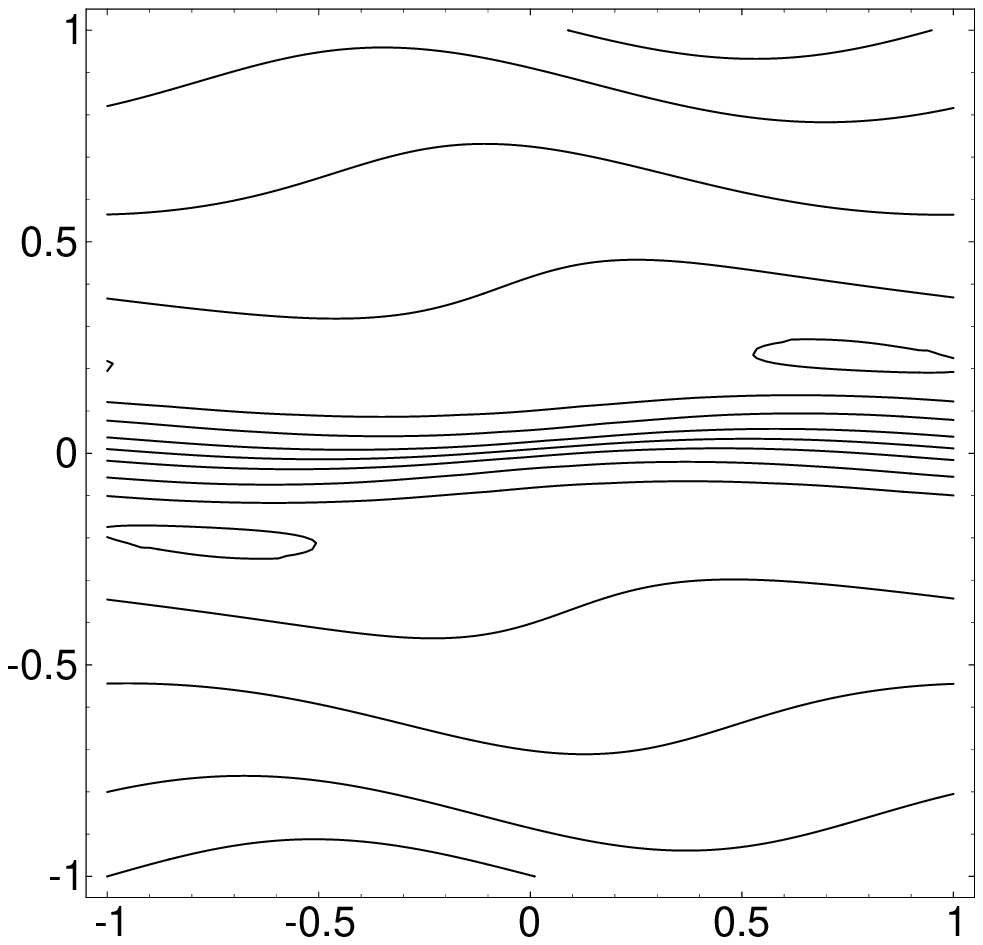}}
	\put(2.2,7.5){\fbox{${\mathcal A}_{I}$ in \% }}
	\put(0.5,-.3){$\varphi_{\mu}/\pi$}
	\put(0,7.3){$ \varphi_{M_1}/\pi$ }
	\put(3.2,6.2){$10$}
\put(3.5,5.5){$20$}
\put(4,4.8){$25$}
\put(6,4.4){$27$}
\put(4.6,4.2){$25$}
\put(5.5,6.6){$0$}
	\put(1.2,3){$-27$}
	\put(3.3,2.6){$-25$}
	\put(4.3,1.8){$-20$}
	\put(5.3,1.2){$-10$}
	\put(2.3,0.7){$0$}
 \end{picture}
\vspace*{.5cm}
\caption{Contour plot for  ${\mathcal A}_{I}$
in the $\varphi_{M_1}$--$\varphi_{\mu}$ plane
for $M_2=400$ GeV, $|\mu|=240$ GeV, 
$\tan \beta=10$, $m_{\tilde \ell_R}=221$ GeV. 
\label{varphases_12}}
\end{minipage}
\hspace*{0.1cm}
				\begin{minipage}{0.45\textwidth}
 \begin{picture}(0,9)(0,0)
	\put(-.5,8.5){\includegraphics{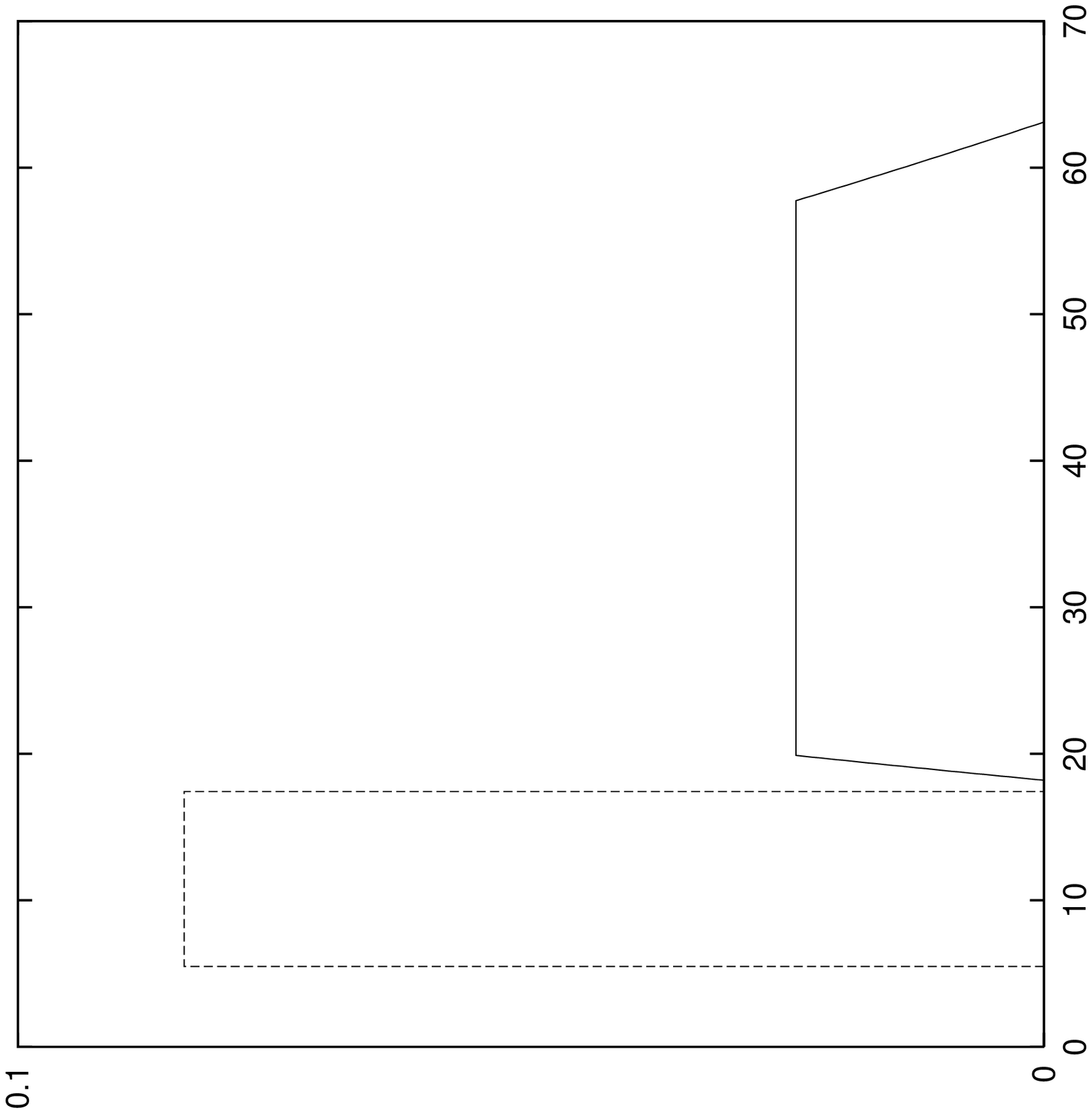}}
	\put(5.5,0.){$E$~/GeV}
	\put(1,8.){$  \frac{1}{\sigma}\frac{d\sigma}{dE}$ }
	\put(1.9,5.5){$ \ell_1 $}
	\put(4.2,2.){$\ell_2 $}
 \end{picture}
\vspace*{.5cm}
\caption{
Energy distributions 
for $\ell_1$ (dashed line) and $\ell_2$ (solid line)
for $M_2=400$ GeV, $|\mu|=240$ GeV, 
$\tan \beta=10$, $m_{\tilde \ell_R}=220$ GeV.
\label{plotedist}}
\end{minipage}
\end{figure}

The sensitivity of the asymmetry ${\mathcal A}_{I}$ on the CP phases 
can be seen from the contour plot in the
$\varphi_{\mu}$--$\varphi_{M_1}$ plane, Fig.~\ref{varphases_12},  
for $|\mu|=240$ GeV and  $M_2=400$.
The asymmetry ${\mathcal A}_{I}$ varies between -27$\%$ and 27$\%$.  
For unpolarized beams this asymmetry would be reduced roughly by a factor 0.33.
It is remarkable that these maximal values are not necessarily obtained 
for maximal  CP phases. In our scenario the asymmetry is 
much more sensitive to variations of the phase $\varphi_{M_1}$ 
around  $0$. On the other hand, the asymmetry is rather insensitive to 
$\varphi_{\mu}$.

The asymmetry ${\mathcal A}_{II}$, which does not require the momentum
identification of the $\tilde{\chi}^0_2$, has a similar dependence 
on the phases as  ${\mathcal A}_{I}$, because both are due to the 
non-vanishing neutralino polarization. 
However, ${\mathcal A}_{II}$ is reduced almost by a factor 3,
because there the CP-violating 
effect from the production is washed out by the kinematics of the 
slepton decay.

The relative statistical error of each asymmetry 
${\mathcal A}$ can be calculated to $\delta {\mathcal A} = 
\Delta {\mathcal A}/{\mathcal A} = S/({\mathcal A} \sqrt{N})$,
with $S$ standard deviations,
assuming a Gaussian distribution of the asymmetry ${\mathcal A}$.
Here, $N={\mathcal L} \sigma$ is the number of events with 
${\mathcal L}$ the total integrated luminosity and 
$\sigma$ the total cross section. Assuming $\delta {\mathcal A}
\approx1$, it follows $S \approx {\mathcal A} \sqrt{N}$.
For example, in order to measure an asymmetry of 5\% with S=2
(confidence level of 95\%), one would need at least $1.5\times 10^3$ events.
This corresponds to a total cross section for reactions
(\ref{production})-(\ref{decay_2}) of 3.1 fb with 
${\mathcal L}=500~{\rm fb}^{-1}$.

The two leptons $\ell_1$ and $\ell_2$ from  the decays
(\ref{decay_1}) and (\ref{decay_2}) have to be distinguished in order to 
measure the asymmetries ${\mathcal A}_{I}$~(\ref{AT1}) and 
${\mathcal A}_{II}$~(\ref{AT2}). 
We show in Fig.~\ref{plotedist} an example of the energy 
distributions of lepton $\ell_1$ (dashed line) and lepton $\ell_2$
(solid line) with $\tan \beta= 10$, $|\mu|=240$ GeV, $M_2=400$ GeV
for $\varphi_{\mu}=0$ and $\varphi_{M_1}=0.5 \pi$. 
The selectron mass is $m_{\tilde{\ell}_R}=220$ GeV, the
LSP mass is $m_{\tilde\chi_1^0}=180$ GeV and the second neutralino mass is 
$m_{\tilde\chi_2^0}=230$ GeV.
Due to the larger mass difference of 40 GeV between $\tilde{\ell}_R$ and 
$\tilde{\chi}^0_1$ compared to the $\tilde{\chi}^0_2$--$\tilde{\ell}_R$ 
mass difference  of 10 GeV,  $\ell_2$ is more energetic than
 $\ell_1$ as the endpoints of the energy distributions 
depend on these mass differences.
The  two distributions shown in Fig.~\ref{plotedist} do not overlap
and one can distinguish between the two leptons event by event.

However, depending on the values of the masses involved, 
there are different types of energy distributions possible. 
The energy distributions also may overlap  if the mass differences
between $\chi^0_1,\tilde{\ell}_R$ and $\tilde{\chi}^0_2 $
are similar. In this case only those leptons can 
be distinguished, whose energies are not both in the overlapping region.
One has to apply cuts which reduce the number of events.

\section{Summary and conclusion
	\label{Summary and conclusion}}

We have considered two CP sensitive triple-product asymmetries in 
neutralino production
$e^+e^- \to\tilde{\chi}^0_i  \tilde{\chi}^0_j$
and the subsequent leptonic two-body decay chain of one  neutralino
$\tilde{\chi}^0_i \to \tilde{\ell} \, \ell$,
$ \tilde{\ell} \to \tilde{\chi}^0_1 \, \ell$ for
$ \ell= e,\mu$. The CP sensitive contributions to the asymmetries
are due to tree level spin effects in the production process 
of an unequal pair of neutralinos. The asymmetries are induced only if 
CP-violating phases of the 
gaugino and higgsino mass parameters $M_1$ and/or $\mu$ are present 
in the neutralino sector of the MSSM.

In a numerical study for $e^+\,e^- \to\tilde{\chi}^0_1 \,
\tilde{\chi}^0_2$ and for the neutralino decay into a right
slepton, we have shown that the asymmetry ${\mathcal A}_{II}$
can go up to 10\% and the asymmetry ${\mathcal A}_{I}$ can be as large
as 25\%. Depending on the MSSM scenario, 
the  asymmetries should be accessible in future  
electron-positron linear collider experiments in the
500 GeV range. Longitudinally polarized electron and positron
beams can enhance both asymmetries considerably.

\section{Acknowledgement}
\vspace*{-.25cm}
This work was supported by the `Fonds zur
F\"orderung der wissenschaftlichen Forschung' (FWF) of Austria, projects
No. P13139-PHY and No. P16592-N02 and by the European Community's
Human Potential Programme under contract HPRN-CT-2000-00149.
This work was also supported by the 'Deutsche Forschungsgemeinschaft'
(DFG) under contract Fr 1064/5-1.

\end{document}